\documentclass[12pt,a4paper]{article}
\usepackage{amsmath}
\usepackage{amssymb}
\usepackage{amsfonts}
\title{Critical percolation and lack of self-averaging in disordered models}
\author{Andrea De Martino\footnote{E-mail address:
dmart@physig.phys.uniroma1.it.}~
and Andrea Giansanti\footnote{E-mail address: giansanti@roma1.infn.it.}\\
{\normalsize \emph{Dipartimento di Fisica, Universit\`a di Roma ``La Sapienza'',}}\\
{\normalsize \emph{Piazzale A. Moro 2, 00185 Roma, Italy}}}
\date{}
\begin{document}
\maketitle
\noindent
{\small \textbf{Abstract -}}
{\small Lack of self-averaging originates in many disordered models from a
fragmentation of the phase space where the sizes of the fragments remain
sample-dependent in the thermodynamic limit. On the basis of new results in
percolation theory, we give here an argument in favour of the conjecture
that critical two dimensional percolation on the square lattice lacks of
self-averaging.}
\vspace{1cm}
\\
The purpose of this note is to discuss a possible relationship between the
arising of non self-averaging effects in disordered models and the occurrence
of a percolative phase transition. More precisely we conjecture as follows:
in two dimensional bond percolation on the square lattice non self-averaging
effects are present at the critical point $p=p_c=\frac{1}{2}$, whereas
above and below the critical probability the model is shown to be
self-averaging. We also suggest that the phenomenon underlying the lack of
self-averaging in many well studied models is a percolation-like phase
transition: whenever non self-averaging effects are present these systems
should be in a critical percolative state. A crucial role for the
justification of these asserts
will be played by the recent progress in two dimensional percolation theory
achieved by Aizenman \cite{aizenman}, Shchur and Kosyakov \cite{shchur},
and Cardy \cite{cardy}. 

The appearance of non self-averaging effects in the low temperature phase
has been the most interesting outcome of the replica approach to spin
glasses \cite{parisi}. It has later been recognized that such
effects are present in a large class of even simpler models
ranging from condensed matter theory to population biology \cite{higgs},
to dynamical systems' theory \cite{derrida::rmm} and mathematics as well
(see Ref. \cite{derrida::overview} for a unifying review of some of them).
In Derrida's words, in the low temperature spin glasses ``phase space
can be thought of as if it was decomposed into infinitely many pure
states $\alpha$, the weights of which remain sample dependent even in the
thermodynamic limit'' \cite{derrida::overview}. This results in non vanishing
sample-to-sample fluctuations of the weights $W_\alpha$ even in the
thermodynamic limit, so that any no matter how large sample is never a
good representative of the whole ensemble. Furthermore, one may find
finite-weighted pure states in each sample, something which sounds strange
since the condition $\sum_\alpha W_\alpha=1$ must always be satisfied.
The same holds for many other models. In some cases, the expression obtained
for the fluctuations of the weights coincides with that obtained for spin
glasses in particular limits (as pointed out in Ref.
\cite{derrida::overview}), so that, again citing Derrida, it looks
like ``the spin glass problem, at least in its mean field version, belongs
to a larger class of problems, and it would be interesting to develop a
more general theory'' \cite{derrida::overview} to treat them.

By now, the standard method to detect non self-averaging effects is that of
studying the quantity
\begin{equation}
Y=\sum_\alpha W_\alpha^2.
\end{equation}
It is possible to show (see Ref. \cite{derrida::overview} for
explicative applications) that if both the ensemble average
$\langle Y\rangle$, and the variance $\textrm{var}(Y)=\langle Y^2\rangle-
\langle Y\rangle^2$ of $Y$ are non zero in the thermodynamic
limit, so that the probability density $\Pi(Y)$ remains ``broad'' when
the system's size goes to infinity, then $Y$ and consequently the weights
$W_\alpha$ are non self-averaging. $\langle Y\rangle$ is the average
of $Y$ over all possible samples, that is over all
possible realizations of disorder, represented by a number of
quenched random variables. 

Broken objects are perhaps the most intuitive and simplest models
showing the same non self-averaging behaviour as spin glasses.
Consider fragmenting a given object of size $1$ into infinite
pieces of sizes $W_\alpha$ according to a given breaking process. 
A sample corresponds to a specific rupture, hence to particular values of a
set of quenched random variables on which the process
depends. For some processes one finds that the sizes of the pieces
lack of self-averaging, that is they remain sample dependent
despite of the fact that the number of pieces is infinite and that
$\sum_\alpha W_\alpha=1$. This is the case of Derrida and Flyvbjerg's
randomly broken object \cite{derrida::rbo},
where the breaking process depends on an infinite number of quenched
random variables. We have recently shown that such a complex procedure is not
necessary for non self-averaging effects to appear. In a geometrically broken
object \cite{adm}, where the breaking process depends on just one random
variable $p$ and the sizes of the pieces form a geometric sequence, the
situation is seemingly less complicated than that of the randomly broken
object, and yet the same non self-averaging effects are present.

Let us now turn to bond percolation and consider a subset of the two
dimensional square lattice $\mathcal{L}_2=\mathbb{Z}^2$ given by a square
$\mathcal{S}(N)$ of $N^2$ points, having $N$ points on each side.
Once the bonds have been assigned with
the usual procedure, namely there is a bond between neighbouring lattice
sites with probability $p$, the square is broken into clusters whose size
is easily identifiable as the number of points forming them, and an isolated
point is regarded as a cluster of size $1$. The sizes normalized with the
total number of points $N^2$ are the weights $W_\alpha$, where $\alpha$ is an
index running from $1$ to the number of clusters formed. Clearly for each
way of assigning the bonds and of forming the clusters one has
$\sum_\alpha W_\alpha=1$. From this viewpoint, percolation theory on
$\mathcal{S}(N)$ studies an object (the square) broken into
pieces (the clusters) according to the well known rule.
The thermodynamic limit is immediately given by the $N\rightarrow\infty$
limit, in which $\mathcal{S}(N)\rightarrow\mathcal{L}_2$. In this limit
percolation theory on the two-dimensional lattice, providing us with bounds
and estimates for the probability that a piece is of a certain size in a
sample produced with acertain value of $p$, is recovered.

A first problem concerns the ensemble in which averages should be
evaluated. We could surely introduce disorder by treating $p$ as a
random variable, namely producing percolation
lattice samples with a value of $p$ chosen from a probability density
$\rho(p)$ on the $[0,1]$ interval, and calculate averages over all possible
choices of $p$, as is done in the geometrically broken object, where the
breaking process depends on just one variable as well. But 
since in bond percolation to a given value of $p$ correspond infinitely
many different breakings, one can define the ensemble of all possible ways
of forming clusters with a given value of $p$ and calculate averages within
that ensemble. This ``fixed $p$'' ensemble average will be denoted
by $\langle\ldots\rangle_p$.

Getting back to bond percolation, let us consider the subcritical phase
(for whatever concerns percolation theory in this paper we refer the
reader to Ref. \cite{grimmett}; all following citations of basic
results in subcritical and supercritical percolation are taken from that
book), where no infinite-sized clusters are present almost surely and the
average cluster size is thus finite (Chapter 5).
It is intuitive that in the thermodynamic limit the probability that two
randomly chosen lattice points belong to same cluster, that is $Y$, goes
to zero, so that the clusters' weights are self-averaging. Nevertheless
the proof that self-averaging holds is quite straightforward. If we denote
by $C_N$ the number of clusters (including isolated points) that are formed
in the square $\mathcal{S}(N)$ and by $\langle n\rangle_p$ the average
number of points in a cluster as a function of $p$, it is possible to
show\footnote{The proof runs as follows: since the weight of a cluster of
size $n$ formed in the square $\mathcal{S}(N)$ is $n/N^2$, one writes
$\langle Y\rangle_p=\lim_{N\rightarrow\infty}\sum_{n=1}^{N^2}(n/N^2)^2 f(n)$,
where $f(n)$ is the average number of clusters of size $n$ formed in the
square. $f(n)$ may be approximated by $C_N P_p(n)$, $P_p(n)$ denoting the
probability that a cluster contains $n$ points. Hence one gets
$\langle Y\rangle_p=\lim_{N\rightarrow\infty}\sum_n \frac{C_N}{N^4}n^2
P_p(n)$, from which formula (\ref{estimate}) soon follows.}
that whenever $p<p_c$
\begin{equation} \label{estimate}
\langle Y\rangle_p\simeq\lim_{N\rightarrow\infty}\frac{1}{N^2}\frac{C_N}{N^2}
\langle n^2\rangle_p .
\end{equation}
From the facts that in the $N\rightarrow\infty$ limit the ratio
$C_N/N^2$ tends to an analytic function $\kappa(p)$ known as the number
of open clusters per vertex (Chapter 4), and that $\langle n^2\rangle_p$ is
finite in the subcritical phase (Lemma 5.89 and in particular its
consequences), follows that $\langle Y\rangle_p\rightarrow 0$
for all $p<p_c$ as $N\rightarrow\infty$. 

For what concerns the supercritical phase, where there is one only infinite
cluster almost surely and the average cluster size diverges (Chapter 6), an
analogous argument yields the conclusion that, if we restrict ourselves to
finite clusters,
\begin{equation}
\langle Y\rangle_p^{\textrm{(finite)}}\rightarrow 0
\end{equation}
when $N\rightarrow\infty$ and $p>p_c$, which means that finite sized clusters
do not contribute to the probability that two randomly chosen lattice points
belong to the same cluster. For the unique infinite cluster we have simply
that $\langle Y\rangle_p^{\textrm{(infinite)}}\geq P_p(\infty)^2$,
where $P_p(\infty)$ is the probability that a cluster has infinite size,
which for $p>p_c$ is finite and positive.

Let's come to the critical point $p=p_c$. Up to recent times it was commonly
believed that there exists exactly one infinite cluster in two dimensional
square lattices at the critical point. This view has been changed by
Aizenman, whose work \cite{aizenman} presents the following main results.
The starting point is the concept of ``spanning cluster'', under which one
understands a cluster that spans a large though finite region of the square
lattice. For example, given the square $\mathcal{S}(N)$, a spanning cluster
is a cluster that transverses it from left to right. Aizenman proved
that when $p=p_c$:
\begin{enumerate}
\item the probability $\Sigma_N(\geq k)$ that $\mathcal{S}(N)$ is transversed
(from left to right) by at least $k$ distinct clusters is strictly positive
for all finite $k$;
\item $\Sigma_N(\geq k)$ satisfies the inequalities
\begin{equation}
A\exp(-\alpha k^2)\leq\Sigma_N(\geq k)\leq\exp(-\beta k^2),
\end{equation}
where $A$, $\alpha$ and $\beta$ are different positive constants.
\end{enumerate}
Furthermore he conjectured that the limit
\begin{equation}
\lim_{k\rightarrow\infty}\lim_{N\rightarrow\infty}
\frac{1}{k^2}\log\Sigma_N(\geq k)
\end{equation}
exists and is finite. Shchur and Kosyakov \cite{shchur} have carried out
numerical studies on the first limit, namely
\begin{equation}
\lim_{N\rightarrow\infty}\Sigma_N(\geq k)=\Sigma(\geq k),
\end{equation}
and have calculated values for the cases $k=2,3$, and an estimate
for the case $k=4$. They found that
\begin{eqnarray}
\Sigma(\geq 2) & \simeq & .00658 ;\nonumber\\
\Sigma(\geq 3) & \simeq & .00000148 ;\\
\Sigma(\geq 4) & \simeq & 10^{-11}.\nonumber
\end{eqnarray}
These results are also in good agreement with those
obtained analytically by Cardy \cite{cardy}, who using methods of conformal
field theory has found the exact $N\rightarrow\infty$ behaviour of the
probability $\Sigma_N(\geq k)$ on a rectangle instead of a square.

These new results suggest further elaboration.
First, in the thermodynamic limit $N\rightarrow\infty$ spanning clusters
become infinite clusters, that is their size diverges. If on one hand we
expect finite clusters to have zero weight in the thermodynamic limit,
on the other hand we expect that infinite clusters may have a finite
weight. If it is so, and if we have such a cluster in a certain sample,
we expect it to give a non zero contribution to the value of $Y$, which
then will be non zero at least for that sample. Now from the recent
results summarized above we know that the number of infinite clusters
that may form at the critical point is sample dependent in the thermodynamic
limit. This holds because the probability $\Sigma(\geq k)$ to find at least
$k$ infinite clusters in $\mathcal{L}_2$ is non zero for all $k$, as
proved numerically by Shchur and Kosyakov
at least for $k=2,3,4$. This means that if we could
produce a large number of samples of the square lattice at $p=p_c$
we would observe samples displaying one infinite cluster, samples
displaying two infinite clusters and so on, so that an histogram
(number of samples with $k$ infinite clusters)vs($k$) would have a finite
variance.
Hence, if infinite clusters at the critical point have finite weight, we
would have that the number of non zero weighted clusters is sample
dependent in the thermodynamic limit. Consequently, also their weights
would remain sample dependent. According to Derrida's
statement on lack of self-averaging as sample dependence in the
thermodynamic limit it is reasonable to conjecture that
if infinite clusters at $p=p_c$ have finite weight then critical two
dimensional percolation on the square lattice lacks of self-averaging.

Proving or disproving the above conjecture might be not too difficult.
Positive results would show that critical percolation and fragmentation
in models of disordered systems have something deep in common. In our
opinion percolation models could well be the starting point to find the
foreseen universality underlying all those disordered models.

\end{document}